# Molecular Autonomous Pathfinder using Deep Reinforcement Learning

Ken-ichi Nomura, Ankit Mishra, Tian Sang, Rajiv K. Kalia, Aiichiro Nakano, and Priya Vashishta

Collaboratory for Advanced Computing and Simulations, University of Southern California, Los Angeles, CA 90089, USA

**Abstract**

Diffusion in solids is a slow process that dictates rate-limiting processes in key chemical reactions. Unlike crystalline solids that offer well-defined diffusion pathways, the lack of similar structural motifs in amorphous or glassy materials poses a great scientific challenge in estimating slow diffusion time. To tackle this problem, we have developed an AI-guided long-time atomistic simulation approach: Molecular Autonomous Pathfinder (MAP) framework based on Deep Reinforcement Learning (RL), where RL agent is trained to uncover energy efficient diffusion pathways. We employ Deep Q-Network architecture with distributed prioritized replay buffer enabling fully online agent training with accelerated experience sampling by an ensemble of asynchronous agents. After training, the agents provide atomistic configurations of diffusion pathways with their energy profile. We use a piecewise Nudged Elastic Band to refine the energy profile of the obtained pathway and corresponding diffusion time on the basis of transition state theory. With MAP, we have successfully identified atomistic mechanisms along molecular diffusion pathways in amorphous silica, with time scales comparable to experiments.

**Main Text**

Water diffusion in silica has been attracting great attention for decades and is central to understanding the mechanical response of silicate materials [1-8]. The presence of water is known to affect the physical and chemical properties of silicates significantly. Stress corrosion cracking (SCC) is an archetypal example where a subcritical crack growth is observed under a moist environment. With moisture, a crack tip of silica glass is found filled with water. These water molecules react with stretched siloxane bonds and break into two silanol groups subjected to tensile loading. A three-stage model is usually used to describe the overall fracture behavior. While simple mechanical argument applies to the $1^{st}$ and $3^{rd}$ stages, the crack growth rate does not depend on applied stress in the $2^{nd}$ stage where molecular diffusion is considered as the rate-limiting step. While a conventional view of SCC is sequential SiO bond breaking by water at the crack tip, recent studies have shown the possibility of fast diffusion pathways mediated by non-bridging oxygens (NBOs) as well as the increased free volume due to stress concentration around the vicinity of crack tip. Water diffusion in silica glass also finds important applications in earth and planetary sciences [9]. To investigate molecular diffusion mechanisms through mantle, silica glass has been used as an experimental platform [2, 3]. Ab-initio molecular dynamics (MD) simulation is an excellent computational tool to study molecular diffusion with atomistic details; however, its computational cost prohibits accessing relevant timescale of molecular diffusion in solids. On the other hand, this slow diffusion process may be seen as a molecule going through a series of energy barriers in a complex energy landscape [10]. To date, several algorithms to enable time-accelerated dynamics have been proposed. However, these methods often require domain specific knowledge or overly simplified reaction coordinates that limit their adaptation to a wide range of material problems. Here, an atomistic simulation framework to explore complex energy landscapes with minimal human intervention and automatically uncover energy efficient diffusion pathways would be ideal.

Here, we propose Molecular Autonomous Pathfinder (MAP) framework combining Deep Reinforcement Learning (DRL) [11-13] and the transition state theory (TST) [14] to investigate water diffusion through silica glass at experimentally-relevant timescale. Reinforcement Learning [15] is a field of machine learning in which an agent interacts with environment to find an optimal policy by maximizing the cumulative reward. DRL extends conventional RL algorithms using advanced Deep Learning techniques and finds diverse applications in robotics, video games, finance, and healthcare, as well as a mission-critical plasma control in the nuclear fusion reactor [16]. DRL has been also applied to molecular simulations, including the optimization of advanced materials synthesis [17, 18], novel drug designs [19, 20], and transition state (TS) search [21]. Deep Q-Network (DQN) is one of the most successful DRL algorithms based on Q-learning [22], which solves the Bellman optimality equation,

$$Q^*(s, a) = \mathbb{E}[r + \gamma \max_{a' \in A} Q^*(s', a')], \qquad (1)$$

where $s$ and $s'$ are the current and next states, $a$ is the action, $\gamma$ is the discount factor, $r$ is the reward, $Q^*$ is the optimal state-action function, and $\mathbb{E}$ denotes an expectation value.

Using DQN, Minh et al. have demonstrated superhuman performance for 49 Atari games by only using pixels and game score as inputs. The Q-function is modeled as Convolutional Neural Network (CNN) to incorporate the complex state definition, *i.e.*, pixel images of the Atari games.

The network parameter is trained by minimizing the temporal-difference (TD) error as loss function given as,

$$L(\theta) = \mathbb{E}_{(s,a,r,s')\sim D}\left[\left(r + \gamma \max_{a'\in A} Q(s', a'; \theta^-) - Q(s, a; \theta)\right)^2\right], \quad (2)$$

where agent's experience $e_t = (s_t, a_t, r_t, s_{t+1})$ is randomly selected from the experience replay buffer $D$. The Q-function is modeled as a three-dimensional CNN. $\theta^-$ and $\theta$ are the parameters of target and behavioral networks to reduce the variance during model training, respectively. Several extensions to the vanilla DQN have been developed to further improve the performance. Hessel et al. proposed Rainbow DQN and achieved roughly 2x improvement for all 57 Atari games [23]. Horgan et al. demonstrated that distributed training could also significantly improve the performance [24]. Their distributed DRL algorithm consists of single Learner process that learns from experiences collected by hundreds of Actor processes. These experiences are prioritized by their TD error value, thereby, Learner can avoid experiences that have already been learned well.

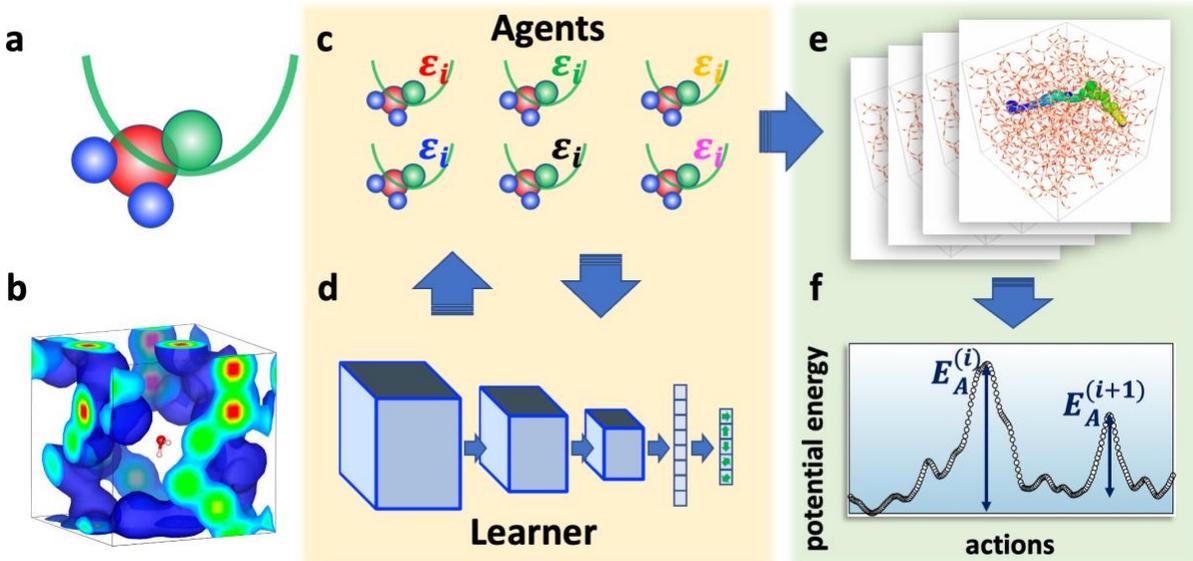

**Fig. 1:** Molecular Autonomous Pathfinder framework. **a,** A schematic of agent (green) navigating a $H_2O$ molecule where red and blue spheres represent O and H atoms, respectively. The agent and the O atom are connected by a harmonic potential interaction indicated by the green curve. **b,** A snapshot of a state. Three-dimensional grid captures the local density distribution of surrounding atoms around the $H_2O$ molecule. The value of the 3D grids is the sum of Gaussian kernel centered at the neighbor atoms. **c** and **d**, A schematic of Agents and Learner processes. Note that each agent has own environment and does not interact other agents. When an agent is instantiated, they are assigned a different $\varepsilon$ values (indicated by $\varepsilon_i$) making some agents favor Q-value predicted by the CNN model (exploitation) and others behave more randomly (exploration). Agents' experiences are sent to Learner process and prioritized with their TD error value. Learner trains three-dimensional CNN that maps the state to the Q-function value for each action by minimizing the TD error. **e** and **f**, Obtained diffusion pathways is further refined using piecewise-NEB algorithm for accurate description of transition states.

With the distributed replay buffer and using 360 Actors, they have achieved approximately 2.3x performance improvement in about the half of state-of-the-art time [24].

**Molecular Autonomous Pathfinder (MAP) Framework**

Fig. 1 schematically presents MAP framework to study long-time water diffusion in silica glass. For the first phase of MAP framework, we use a scalable DRL algorithm to train autonomous molecular agents to find energy-efficient diffusion pathways. While offline RL training takes advantage of precomputed training data, it suffers from the distributional shift problem, which is one of the central challenges in offline reinforcement learning [4]. On the other hand, the application of online learning is severely limited by its sample inefficiency. To realize a generic framework that is applicable to a wide class of materials, MAP employs online training to avoid the distributional shift with linear-scaling sampling of experiences by distributed asynchronous agents. See Fig. S1 in Supplementary Information. In the second phase, the obtained diffusion pathways are divided into mutually exclusive segments that contain two potential energy minima with one energy maximum in-between. We apply a piecewise Nudged Elastic Band (pNEB) method on each segment to refine the energy barriers. Based on transition state theory [25], the associated time of each segment is estimated as,

$$T_m = \Sigma_{i \in \{1,\ldots,NE_A\}} \frac{\hbar}{k_B T} \exp\left(\frac{E_A^{(i)}}{k_B T}\right), \qquad (3)$$

where $\hbar$ is the reduced Plank constant, $k_B$ is Boltzmann constant, $T$ is temperature, $NE_A$ is the total number of energy barriers, and $E_A^{(i)}$ is $i$-th energy barrier along the pathway respectively. Since each NEB calculation can be done in parallel, our MAP framework efficiently evaluates the total diffusion time on massively parallel supercomputers.

**Environment:** The environment is modeled by Reactive Molecular Dynamics (RMD) simulation. To incorporate the energetics of bond breaking and formation during agent training, we employ quantum-mechanically validated ReaxFF [26, 27] force field and a scalable MD software RXMD [28]. Silica glass structure was created by melt-quench method [29]. The system dimensions are (30.48 Å)$^3$. The total number of atoms are 1,593 including one $H_2O$ molecule navigated by the RL agent. Throughout the training, the system is thermalized at 10K with the canonical (NVT) ensemble with a timestep of 0.5 fs.

**Agent:** an agent is defined as a harmonic interaction function that is bound to an atom to facilitate molecular diffusion through silica glass; Fig. 1a. The center of the harmonic potential, *i.e.* the agent position, is bound to the O atom of $H_2O$ molecule and updated following the agent's action. We employ a set of discrete actions that consists of five displacement vectors $\vec{a} = [(1,0,0), (0,1,0), (0,0,1), (0,-1,0), (0,0,-1)]$ to avoid the agent oscillation problem [30]. State, $s$, is defined as a three-dimensional grid that approximates the distribution of neighbor atoms around the agent. From each neighbor atom, Gaussian kernel is used to represent their contribution to the local density. Based on the predicted Q-function value given a state $s$, agent updates its position $\mathbf{r}_{agent}$ by a chosen vector multiplied by a displacement magnitude $\delta$.

$$\mathbf{r}_{agent} = \mathbf{r}_{agent} + \delta \operatorname*{argmax}_{a} Q(s, a) \tag{4}$$

To facilitate agents' exploration, we use $\varepsilon$-greedy policy where an action is randomly chosen with the probability of $\varepsilon$ value regardless of the Q-value. For the distributed training, each agent is assigned a different randomness factor $\epsilon = (0, \epsilon_{max})$ to control their extent of exploration; Fig. 1c. After an action has been taken, the $H_2O$ molecule and silica system are relaxed by a short RMD simulation.

**Learner:** The main tasks of Learner process are to organize agents' experiences by their TD error, update model parameters of the Q-function, and synchronize the updated model with agents. Whenever Learner receives agent's experience, the Learner computes the TD error and stores the tuple of experience and TD error in the replay buffer. To compute the loss function Eq. (2), Learner randomly samples batch-sized tuples from the reply buffer with a probability proportional to the TD error. A large TD error value indicates an unlearned situation for the model. The prioritized reply buffer is particularly useful when many agents tend to accumulate experiences around the initial starting location [24].

**Reward:** We use five reward functions in total to train the agents. The function $R_{position}$ gives reward based on the distance between a predefined goal and the location of the agent. In this study, we use the *x*-coordinate of the agent as the reward value. This monotonically increasing function serves as the baseline of the overall reward structure. $R_{energy}$ rewards an agent if the current state has a lower potential energy than a reference potential energy. To minimize the noise in the reference energy, we use the mean of the potential energy over the last one hundred MD steps. $R_{density}$ penalizes the agent, *i.e.,* gives a negative reward, if the distance between the agent and surrounding atoms becomes too small. Similarly, we also apply a penalty $R_{distance}$ if the distance between the agent and water molecule becomes greater than a prescribed threshold. Technically an agent may earn rewards by making actions with positive rewards without moving from the same location. To achieve efficient learning of the environment, a common practice is to apply a time penalty, $R_{time}$, with which an agent receives a negative reward whenever a new action is made. Total reward $R_{total}$ is the sum of the five rewards with weighting perfectors that are tuned as hyperparameters.

Given the reward structure, we have investigated the training efficiency as a function of the number of agents, $N$. As the measure of the agent's performance, we record the agent's $x$-coordinate and total reward at the end of each epoch. Fig. 2 shows a typical agent's performance with $N = 1$, 4, and 16 during 48 hours of training. Overall, increasing the number of agents results in better performance for both the agent's final coordinate and the total reward. With $N = 1$, the total reward is kept low although the final position consistently increases. With $N = 4$, both metrics increase at the beginning however the increase rate slows down after 10,000 steps. With $N = 16$, the final agent position increases further with a smaller deviation. While the final position is comparable with $N = 4$ and 16 up to 10,000 steps, the total reward with $N = 16$ is noticeably better, approximately 2 times greater than the one with $N = 4$.

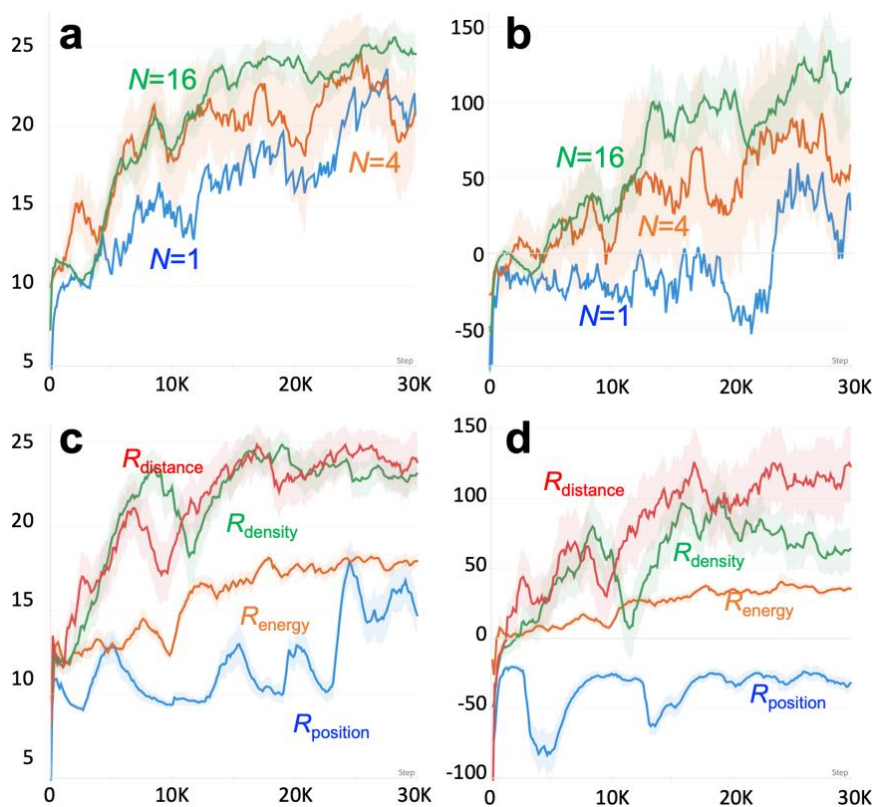

**Fig. 2**: **a** and **b,** Training performance as a function of the number of agents. **a** and **b** show the agent's final position and the total reward with $N = 1$, 4, and 16, respectively. For the training with N > 1, the solid line shows the mean of all agents' peformance and the shade represents their standard deviation. Overall, the training performance improves by increasing $N$. While the final agent position is comparable with $N = 4$ and 16 up to 10,000 steps, the total reward with $N = 16$ is approximately 2 times greater than the one with $N = 4$, indicating that more efficient pathways have been discovered with $N = 16$. **c** and **d**; An ablation study on the reward functions where the agent's final position and total reward are monitored with one of the four rewards ($R_{position}$, $R_{energy}$, $R_{density}$, and $R_{distance}$) are turned off. All results are obtained with $N = 16$. $R_{position}$ serves the baseline reward while $R_{energy}$ helps agents find diffusion pathways with greater total rewards.

Next, we have performed an ablation study with one of the reward functions turned off. See Fig. 2c and d. We observe that $R_{position}$ serves as the baseline reward for an agent to learn the proper direction to move. With $R_{position}$ off, the agent's final position remains around 13 Å (out of 30.48 Å) and the total reward remains negative. With $R_{energy}$ turned off, the agent also appears to struggle, resulting in a suboptimal final position around 18Å and the value of total reward is about 30, respectively. On the other hand, a decent training performance is obtained with either $R_{density}$ or $R_{distance}$ turned off, signifying the important contribution of $R_{position}$ and $R_{energy}$ for agents to learn the environment.

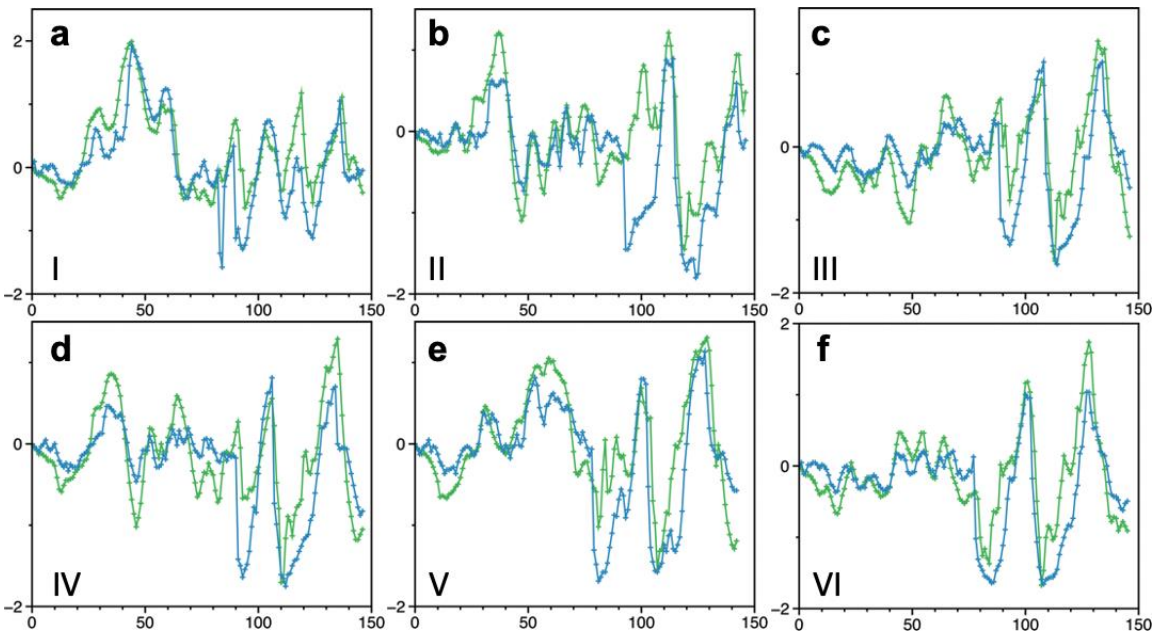

**Fig. 3**: **a-f,** Potential energy profile before (green) and after (blue) pNEB on the last six energy efficient pathways using $N = 16$ after 48 hours of training. The vertical axis is the potential energy of the system in eV unit and the horizontal axis is the number of actions. The cumulated number of actions of the episode I-VI is 31,068, 30,752, 28,243, 27,835, 24,020, 23,989, respectively. The initial, final and transition states in the original energy profile are identified by SciPy signal processing library. The five episodes except the episode I have converged to a similar profile that consists of an initially flat region followed by two consecutive energy barriers approximately around $100^{th}$ and $120^{th}$ action. The episode I shows an additional peak around $120^{th}$ action.

Once sufficient pathways have been sampled, we apply pNEB to refine the description of transition states and use Eq. 3 to estimate the diffusion time. Fig. 3 shows the potential energy profile of the top six episodes (episodes I to VI) ranked by their diffusion time after 48 hours of training using $N = 16$. The original energy profile obtained by RL agent is divided into several segments, each of which contains one initial, final and transition state. Subsequently, pNEB is simultaneously performed on each segment in the energy profile. After the pNEB calculation, the five episodes (II-VI) have converged to a similar energy profile, in which a rather flat region continues up to around the 90$^{th}$ action, followed by two noticeable energy barriers at 100$^{th}$ and 120$^{th}$ actions. In addition to the existing two barriers, an additional TS with a relatively small activation energy (~ 0.8 eV) is observed at 117$^{th}$ action in the episode I.

To obtain atomistic insights along the diffusion pathway, Fig. 4 presents a series of snapshots of the transition states of the episode I shown in Fig. 3a. See S1.mov in Supplementary Materials for a complete trajectory. In the episode, the H$_2$O molecule is first weakly absorbed by a siloxane bond [31] and detached the site after a few actions. After 80$^{th}$ action, the H$_2$O molecule is moved to an undercoordinated Si site, which results in the reduction of potential energy by close to 1 eV. Soon after, the H$_2$O molecule hops to another undercoordinated Si site by a concerted switching of neighboring Si atoms; Fig. 4c. Finally, the H$_2$O molecule detaches from the silicon, labeled as Si3 in Fig. 4d. In the episode I, however, the hopping between the two Si atoms occurs in two steps, shown in Fig. 4e and f. Instead of the direct hop, the H$_2$O molecule approaches the Si3 moving around Si2, accumulating the strain energy by distorting the silica network. The acquired strain energy facilitates the desorption of the H$_2$O molecule from the Si2 site, reducing the overall lower energy barrier. Consequently, the estimated diffusion time by the two-step hopping mechanism is reduced by a factor of 15, namely 0.124 days vs. 1.93 days to travel 3 nm at 350 °C in the episodes I and II, respectively. The obtained timescale is comparable to experimental observations [3, 5], providing an atomistic insight for the experiments.

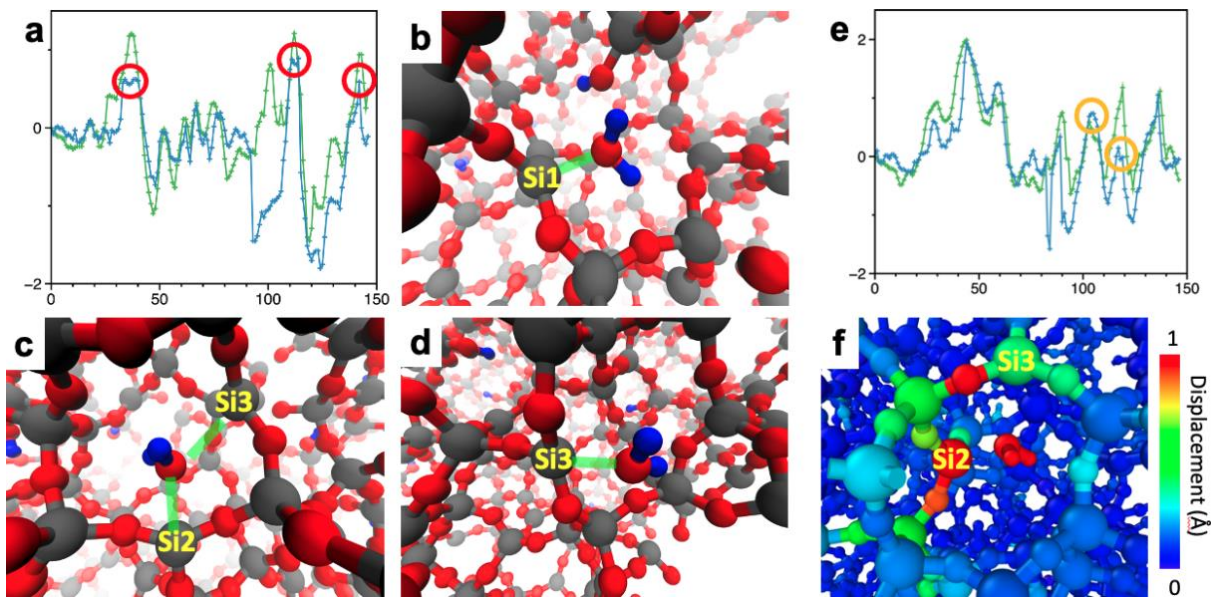

**Fig. 4**: **a**, Potential energy profile before (green) and after (blue) pNEB and three TSs (red circles) in the episode II. **b-d,** the atomic configurations of the three TSs in **a**. Green lines in **b-d** indicate the closest Si from the O in the RL-guided H$_2$O molecule. **e**, Potential energy profile of the episode I before (green) and after (blue) pNEB along with the two TSs (orange circles) during the H$_2$O molecule hopping from Si2 and Si3. **f**, The atomic configuration of the second TS color-coded by the atomic displacement from their initial position. A large displacement (over 1Å) on Si2 indicate the stored mechanical strain energy in the TS.

**Conclusions**

We have developed a scalable AI-guided framework combining DRL and TST to study the long-time water diffusion process in silica glass. MAP framework based on DQN architecture realizes autonomous agents to explore the complex energy landscape of the silica glass network to obtain energy-efficient diffusion pathways. Obtained energy profiles are further refined using the piecewise NEB to efficiently translate to the overall diffusion time. The algorithmic design within MAP framework particularly focuses on its transferability to tackle a wide class of material problems; for example, the use of online learning to eliminate the necessity of generating training datasets that often requires expert domain knowledge, and the highly efficient sampling of agent experiences by taking advantage of advanced computing architectures. Not only the inorganic silica glass presented in this study, MAP framework has been successfully applied to organic polymer systems [32], providing a novel approach to investigate long-time diffusion mechanisms with atomistic-level insights.

**Acknowledgment**

Research supported by the U.S. Department of Energy, Office of Basic Energy Sciences, Division of Materials Sciences and Engineering, Neutron Scattering and Instrumentation Sciences program under Award DE‑SC0023146.